\documentclass[10pt,twocolumn,letterpaper]{article}

\usepackage[pagenumbers]{cvpr} 

\usepackage{multirow}
\usepackage{tabularx}
\usepackage{bm}
\usepackage{natbib}

\usepackage{listings}
\usepackage{float}
\usepackage{xcolor,colortbl}
\usepackage[accsupp]{axessibility}

\usepackage{inconsolata}

\usepackage{xcolor}

\definecolor{codegreen}{rgb}{0,0.6,0}
\definecolor{codegray}{rgb}{0.5,0.5,0.5}
\definecolor{codepurple}{rgb}{0.58,0,0.82}
\definecolor{backcolour}{rgb}{0.95,0.95,0.92}
\definecolor{white}{rgb}{1.0,1.0,1.0}

\definecolor{cvprblue}{rgb}{0.21,0.49,0.74}
\usepackage[pagebackref,breaklinks,colorlinks,allcolors=cvprblue]{hyperref}

\def\paperID{10211} 
\def\confName{CVPR}
\def\confYear{2025}

\title{Real-time High-fidelity Gaussian Human Avatars with \\ Position-based Interpolation of Spatially Distributed MLPs}

\author{Youyi Zhan$^1$ \quad Tianjia Shao$^1$\footnotemark[1] \quad Yin Yang$^2$\quad Kun Zhou$^1$\\
$^1$State Key Lab of CAD\&CG, Zhejiang University \quad $^2$University of Utah \\
}

\begin{document}

\twocolumn[{%
\renewcommand\twocolumn[1][]{#1}%
\maketitle
\thispagestyle{empty}
\vspace{-1cm}
\begin{center}
     \includegraphics[width= 0.99\linewidth]{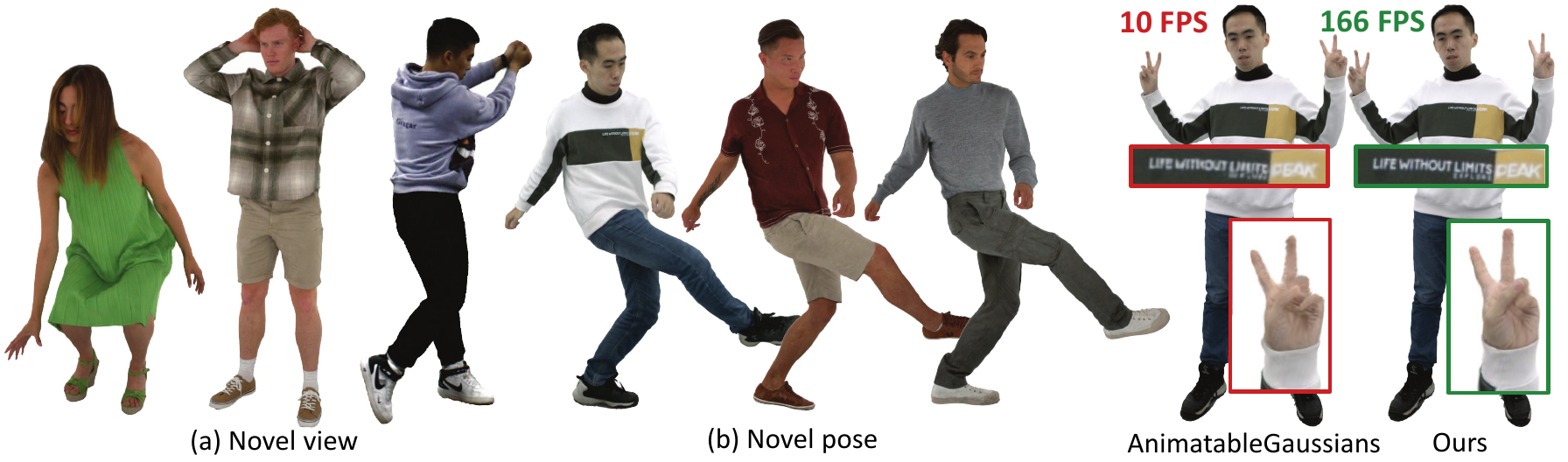}
     \captionof{figure}{Our method can model high-fidelity human avatars that can be animated under novel poses and rendered in real-time. Compared to the state-of-the-art method AnimatableGaussians~\cite{li2024animatable}, our approach can recover finer details while achieving significantly faster rendering speed (166 fps) under novel views and novel poses. 
        }
    \label{fig:teaser}
\end{center}
}]

\maketitle

\renewcommand{\thefootnote}{\fnsymbol{footnote}}
\footnotetext[1]{Corresponding author (tjshao@zju.edu.cn)}

\begin{abstract}
Many works have succeeded in reconstructing Gaussian human avatars from multi-view videos. 
However, they either struggle to capture pose-dependent appearance details with a single MLP, or rely on a computationally intensive neural network to reconstruct high-fidelity appearance but with rendering performance degraded to non-real-time.
We propose a novel Gaussian human avatar representation that can reconstruct high-fidelity pose-dependence appearance with details and meanwhile can be rendered in real time. Our Gaussian avatar is empowered by spatially distributed MLPs which are explicitly located on different positions on human body. The parameters stored in each Gaussian are obtained by interpolating from the outputs of its nearby MLPs based on their distances. To avoid undesired smooth Gaussian property changing during interpolation, for each Gaussian we define a set of Gaussian offset basis, and a linear combination of basis represents the Gaussian property offsets relative to the neutral properties. Then we propose to let the MLPs output a set of coefficients corresponding to the basis. In this way, although Gaussian coefficients are derived from interpolation and change smoothly, the Gaussian offset basis is learned freely without constraints. The smoothly varying coefficients combined with freely learned basis can still produce distinctly different Gaussian property offsets, allowing the ability to learn high-frequency spatial signals. We further use control points to constrain the Gaussians distributed on a surface layer rather than allowing them to be irregularly distributed inside the body, to help the human avatar generalize better when animated under novel poses. Compared to the state-of-the-art method, our method achieves better appearance quality with finer details while the rendering speed is significantly faster under novel views and novel poses.
\end{abstract}

\vspace{-5mm}
\section{Introduction}
\label{sec:intro}

Digital human avatars have widespread applications in the fields like virtual reality and visual content creation. The reconstruction of human avatars from multi-view videos has been extensively studied, and remarkable progress is being made with the radiance field representation (i.e., NeRF~\cite{mildenhall2020nerf} and 3D Gaussians~\cite{kerbl20233d}). Specially, with the fast training and rendering speed, many works have succeeded in using 3D Gaussians to reconstruct digital human avatars with high quality pose-dependent appearances.

Among those works, a common way is to use a single multilayer perceptron (MLP) network~\cite{ye2023animatable,qian20243dgs,moreau2024human} , which takes human pose along with Gaussian position or learnable code as input and outputs Gaussian property offsets for each Gaussian. Though these methods can achieve real-time rendering performance (i.e., 30-60 fps), due to limited learning capacity, they fail to capture high-frequency details. To reconstruct high-fidelity detailed appearances, the state-of-the-art work AnimatableGaussians~\cite{li2024animatable} proposes to use more powerful network StyleUNet~\cite{wang2023styleavatar} to predict Gaussian property maps with the posed position map as input. AnimatableGaussians~\cite{li2024animatable} can reconstruct the highest-quality human avatar among existing methods. However, due to heavy computational burden brought by StyleUNet, the rendering performance is not real time (i.e., around 10 fps). 

Our goal is to reconstruct high-fidelity detailed Gaussian human avatars, which can be rendered in real time under novel views and novel poses. To this end, we propose a new Gaussian human avatar representation, empowered by spatially distributed MLPs. Different from the single MLP which takes the position (or learnable code) and pose as input, the spatially distributed MLPs are explicitly located on different anchor positions on human body, and their input is only the human pose. The parameters stored in each Gaussian are obtained by interpolating from the outputs of its nearby MLPs based on distances. In this way, each MLP is only responsible for learning the the human appearance of its local region, reducing the learning burden and enhancing the capability of capturing high-frequency details. Besides, with the position-based interpolation, we don't need to send every Gaussian position to the MLP anymore, avoiding going through the MLP too many times. For a pose, each MLP only needs to be computed once, so the rendering performance can be significantly accelerated, reaching 166 fps even with 200K Gaussians under novel poses.

However, what to interpolate is not trivial. If we let the MLPs output Gaussian property offsets as the single MLP in previous methods, and perform interpolation on the Gaussian property offsets, we will obtain smoothly changing Gaussian property offsets across the body, which is undesirable and will produce artifacts (see~\cref{fig:ablationinterprop} for example). To this end, for each Gaussian, we define the neutral properties representing the mean appearance, and a set of Gaussian offset basis. A linear combination of the basis represents the Gaussian property offsets relative to the neutral properties. Then we propose to let the MLPs output a set of coefficients, and the Gaussian coefficients for the offset basis are obtained by interpolation. The key insight of our design is, although Gaussian coefficients are derived from interpolation and change smoothly, the Gaussian offset basis is learned freely without constraints. Therefore, the smoothly varying coefficients combined with freely learned basis can still produce distinctly different Gaussian property offsets. This allows the Gaussians to have the ability to learn high-frequency spatial signals.

Furthermore, we propose to use control points to constrain the Gaussians distributed on a surface layer rather than allowing them to be irregularly distributed inside the body. Specifically, the Gaussian position offset from the neutral position is not optimized freely, but is interpolated from the offsets of sampled control points on the body. By constraining the neighboring control points to have similar position offsets, the Gaussians among these control points are also constrained to move in the same direction without undesired moving inside. This design helps the human avatar generalize better and eliminates artifacts when animated under novel poses.

Experiments demonstrate that our method can reconstruct high-fidelity appearance with high-frequency details of human avatars. Compared to the state-of-the-art method, our method achieves better appearance quality with finer details while the rendering speed is significantly faster under novel views and novel poses (166 fps versus 10 fps).

\section{Related Work}

\noindent \textbf{Mesh based Human Avatar.}
Using mesh with texture is the most common approach to model human avatars. Through learning from video, many methods reconstruct geometry for individuals and apply textures to obtain appearance. \cite{collet2015high,bagautdinov2021driving,xiang2021modeling,xiang2022dressing} use dense camera arrays to reconstruct geometry. \cite{tong2012scanning,bogo2015detailed} further use depth cameras to assist in geometry reconstruction. To obtain appearances under different poses, \cite{habermann2019livecap,habermann2021real,ma2021pixel,xiang2023drivable,xiang2021modeling,xiang2022dressing} use neural networks to output textures for different poses. \cite{xiang2021modeling,xiang2022dressing,xiang2023drivable} further model clothing as a separate layer, achieving realistic results with garment dynamics.

\noindent \textbf{Neural Rendering for Human Avatar.} 
In recent years, neural radiance field (NeRF~\cite{mildenhall2020nerf}) has been widely used for human avatar reconstruction. Many methods~\cite{peng2021neural,peng2021animatable, liu2021neural,zheng2022structured,li2022tava,weng2022humannerf,jiang2022neuman,yu2023monohuman,li2023posevocab,liu2024texvocab} render the human avatar by inverse LBS and obtaining attributes (like color and density) for volume rendering, and achieve good results. ARAH~\cite{wang2022arah} and Vid2avatar~\cite{guo2023vid2avatar} further use signed distance function (SDF) to represent human geometry. However, due to the requirement of multiple sampling, these methods are inefficient, resulting in several seconds to render a single image.

Some works focus on accelerating the above rendering process to achieve faster training and rendering speeds. InstantNVR~\cite{geng2023learning} and InstantAvatar~\cite{jiang2023instantavatar} use iNGP network~\cite{mueller2022instant} to speed up training. However, these methods fail to capture human details, resulting in less realistic rendering. AvatarRex~\cite{zheng2023avatarrex}, Deliffas~\cite{kwon2024deliffas}, UV Volume~\cite{chen2023uv} and RAM-Avatar~\cite{deng2024ram} achieve high-quality appearances at real-time speeds. AvatarRex~\cite{zheng2023avatarrex} is a method for full body avatar with face, body and hands, where SLRF~\cite{zheng2022structured} and dynamic feature patches are used to model the body geometry and color. Despite some breakthroughs in speed, these methods still rely on extensive network computation to obtain appearances, limiting their speed to around 10-25 fps.

\noindent \textbf{3DGS based Human Avatar.}
3D Gaussian Splatting (3DGS~\cite{kerbl20233d}) provides a new paradigm for scene reconstruction. Many works have successfully use 3DGS for modeling human avatars. \citet{kwon2025generalizable} and GPS-Gaussian~\cite{zheng2024gps} are able to reconstruct the avatar from multi-view cameras, but their avatar cannot be driven by novel poses. SplattingAvatar~\cite{shao2024splattingavatar},  HAHA~\cite{svitov2024haha}, GomAvatar~\cite{wen2024gomavatar}, \citet{moon2024expressive}, EVA~\cite{hu2024expressive}, Gauhuman~\cite{hu2024gauhuman}, GART~\cite{lei2024gart}, iHuman~\cite{paudel2024ihuman}, HUGS~\cite{kocabas2024hugs} and SplatArmor~\cite{jena2023splatarmor} propose to reconstruct human avatars from monocular video. However, they cannot model pose-dependent appearance. 3DGS-Avatar~\cite{qian20243dgs}, \citet{ye2023animatable}, \citet{moreau2024human} propose to use a single MLP to output Gaussian property offsets under different poses. They typically use pose along with positions~\cite{qian20243dgs} or per-Gaussian learnable codes~\cite{ye2023animatable,moreau2024human} as MLP inputs to predict Gaussian properties. However, these methods fail to capture high-frequency details due to limited learning capacity. Ash~\cite{pang2024ash}, UV Gaussians~\cite{jiang2024uv}, MeshAvatar~\cite{chen2024meshavatar} proposes to use convolutional neural network (CNN) to output Gaussian property maps, but these works still fail to reconstruct realistic human avatars. Both AnimatableGaussians~\cite{li2024animatable} and DEGAS~\cite{shao2024degas} use large CNN to learn appearance and achieve high-quality rendering. However, it is slow to go through their network, as their large CNNs involve substantial computation, limiting the rendering speed.  Instead, our approach doesn't involve heavy neural network computations and can render high-quality avatar at faster speed.

\noindent \textbf{Discussion of Highly Related Works.}
For the MLPs, both SLRF~\cite{zheng2022structured} and our method employ spatially distributed MLPs to learn local appearance. However, SLRF takes position encoding as MLP input, which requires a huge amount of position queries to the MLPs, largely decreasing the inference speed, while our MLPs only take the human pose as input, significantly reducing the computational burden. For the linear basis, \citet{gao2022reconstructing} and \citet{ma20243d} learn the blendshape basis while the coefficients are from the FLAME~\cite{li2017learning} model and not learnable. Our method jointly learns the coefficients and basis.

\begin{figure*}[t]
  \begin{center}
    \includegraphics[width=\textwidth]{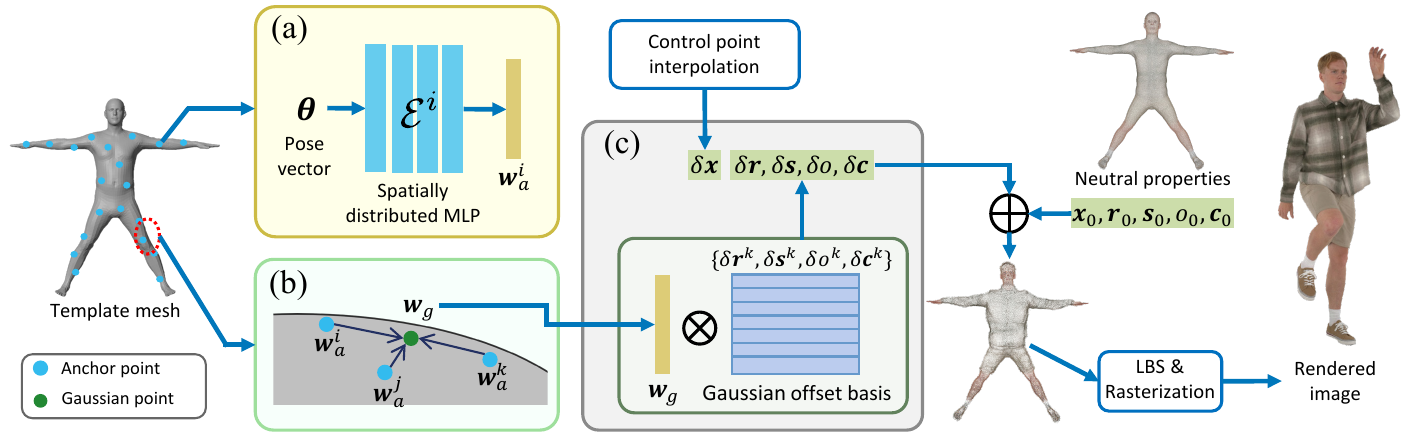}
  \end{center}
  \vspace{-5mm}
  \caption{Pipeline overview. (a) We define the spatially distributed MLPs on anchor points, which are uniformly sampled on the template mesh. Each MLP takes the pose $\bm{\theta}$ as input and outputs the anchor coefficients $\mathbf{w}_a$. (b) The Gaussian coefficients  $\mathbf{w}_g$ are interpolated from the coefficients of three nearest anchor points. (c) The Gaussian property offsets are obtained by linearly combining Gaussian offset basis using Gaussian coefficients. Then the neutral Gaussian properties are added with Gaussian property offsets to model the human appearance under pose $\bm{\theta}$. Finally the Gaussians are transformed to the pose $\bm{\theta}$ and rasterized to produce high-fidelity images. Note that the Gaussian position offset $\delta \mathbf{x}$ is obtained through control point interpolation, which is illustrated in \cref{sec:anchorpoints}.
  }
  \label{fig:pipeline}
  \vspace{-3mm}
\end{figure*}

\section{Method}

Our approach takes the multi-view videos of a person as input. Following previous methods~\cite{li2024animatable,zheng2023avatarrex,isik2023humanrf}, we extract the foreground human mask, and register the SMPL-X~\cite{pavlakos2019expressive} model for each frame to obtain the 3D human pose. We also use the method of AnimatableGaussians~\cite{li2024animatable} to obtain a canonical template mesh.
Our goal is to reconstruct a human avatar, which has pose-dependent high-fidelity detailed appearances under novel views and novel poses, and meanwhile can be rendered in real-time. We first introduce our Gaussian avatar representation with spatially distributed MLPs (\cref{sec:avatarrep}). Then we propose to use control points to obtain per Gaussian position offset, so that Gaussians can be constrained on a surface layer (\cref{sec:anchorpoints}). Finally, we describe the training and testing process, as well as implementation details (\cref{sec:trainandtest}).

\subsection{Gaussian Avatar with Spatially Distributed MLPs}
\label{sec:avatarrep}

Our avatar is composed of $N$ Gaussians and $F$ spatially distributed MLPs.  Each Gaussian has a set of neutral properties $ \Lambda_0$, including rotation $\mathbf{r}_0$, scale $\mathbf{s}_0$, opacity $o_0$, SH coefficients $\mathbf{c}_0$, and position $\mathbf{x}_0$. The neutral properties represent the mean human appearance across the video frames. We also define a set of Gaussian offset basis $\delta \Lambda^k = \{ \delta \mathbf{r}^k, \delta \mathbf{s}^k, \delta o^k, \delta \mathbf{c}^k\}, \ k \in [1,B]$ for each Gaussian. A linear combination of the basis represents a Gaussian property change relative to the neutral properties.

The spatially distributed MLPs are located on $F$ anchor points $\{ \mathbf{x}_a^j \}_{j\in [1,F]}$ uniformly sampled on the template mesh. Each MLP is only responsible for learning the local appearance change around the anchor point. The MLP takes the human pose vector $\bm{\theta}$ as input, and outputs the anchor coefficients $\mathbf{w}_a^j$ on each anchor point,

\begin{equation}
\mathbf{w}_a^j = \mathcal{E}^j(\bm{\theta}),
\label{eqn:anchorw}
\end{equation}
where $\mathcal{E}^j$ is the spatially distributed MLP located on the $j$th anchor point. Based on the anchor coefficients, we obtain the Gaussian coefficients $\mathbf{w}_g$ for the offset basis on each Gaussian by interpolating from the nearest three anchor points, 
\begin{equation}
\mathbf{w}_g = \frac{ \sum_j  \gamma(\mathbf{x}_0, \mathbf{x}_a^j) \cdot \mathbf{w}_a^j }{ \sum_j   \gamma(\mathbf{x}_0, \mathbf{x}_a^j)  },
\label{eqn:interpwght}
\end{equation}
where $\gamma(\mathbf{x},\mathbf{y}) = 1 / \| \mathbf{x} - \mathbf{y} \|_2$ is the reciprocal of the distance between two points. $j$ is the index of three nearest anchor points.

We linearly combine the Gaussian offset basis using the Gaussian coefficients to obtain the property offset under pose $\bm{\theta}$ for each Gaussian, and the Gaussian properties are obtained by adding the property offset to the neutral Gaussian,
\begin{equation}
\begin{aligned}
\delta \Lambda &= \sum_{k=1}^{B} \mathbf{w}_g[k] \cdot \delta \Lambda^k \\
\Lambda &= \Lambda_0 + \delta \Lambda. 
\end{aligned}
\label{eqn:propblend}
\end{equation}
Please note the Gaussian position is also computed as $\mathbf{x} = \mathbf{x}_0 + \delta \mathbf{x}$, nevertheless the position offset $\delta \mathbf{x}$ is actually interpolated from the position offsets of control points, which is detailed in \cref{sec:anchorpoints}.

Afterwards, the Gaussians are transformed from the canonical space to the pose $\bm{\theta}$ using linear blend skinning (LBS). The transformed Gaussians are finally rasterized to produce high-fidelity detailed human images.

\begin{figure}[t]
  \begin{center}
    \includegraphics[width=0.95\columnwidth]{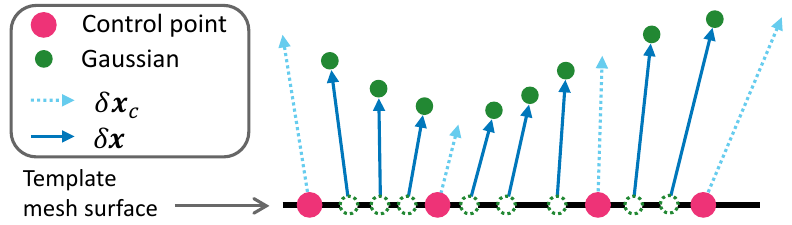}
  \end{center}
  \vspace{-5mm}
  \caption{Illustration of the control point. The Gaussian position offset $\delta \mathbf{x}$ is interpolated from the position offsets of nearby control points $\delta \mathbf{x}_c$. 
  }
  \label{fig:anchorpoint}
  \vspace{-5mm}
\end{figure}

\subsection{Control Point}
\label{sec:anchorpoints}

We do not calculate the position offset using \cref{eqn:propblend} because this would allow each Gaussian to move freely in space during optimization. Irregularly distributed Gaussians in space can produce non-neglectable artifacts (see \cref{fig:ablationanchorpoint}  ``w/o control point"). 
Therefore, we need to constrain the Gaussians distributed on a surface layer rather than allowing them to be irregularly distributed inside the body.

A straightforward solution is to use smoothness loss to constrain neighboring Gaussians to have similar position offsets from the neutral positions. However, we find such design can only ensure very local position smoothness and cannot prevent Gaussians from moving inside the body, resulting in suboptimal results (see \cref{fig:ablationanchorpoint} ``w/o control point (w/ smooth)" and supplementary video). To this end, we propose to use sampled control points to yield similar position offsets across larger areas, and interpolate the position offsets on control points to generate the Gaussian position offsets.

Specifically, we uniformly sample $C$ control points $\{ \mathbf{x}_c^i \}_{i\in [1,C]}$ on the template mesh. 
Each control point has its neutral position offset $\delta \mathbf{x}_{c0}$, as well as a set of position offset basis $\{ \delta \mathbf{x}_{cb}^k \}_{k\in [1,B]}$. We compute the position offset $\delta \mathbf{x}_c$ for each control point by using the control point coefficients $\mathbf{w}_c$ to combine the offset basis and adding the neutral offset,
\begin{equation}
\delta \mathbf{x}_c = \delta \mathbf{x}_{c0} + \sum_{k=1}^B \mathbf{w}_c[k] \cdot \delta \mathbf{x}_{cb}^k.
\end{equation}
$\mathbf{w}_c$ is computed by interpolating the anchor coefficients of position offsets, similar to \cref{eqn:interpwght}. Note the anchor coefficients of position offsets are simultaneously outputted from the position-ware MLP $\mathcal{E}^j(\bm{\theta})$ in \cref{eqn:anchorw}

Then, the Gaussian position offset $\delta \mathbf{x}$ is obtained by interpolating the position offsets of its three nearest control points,
\begin{equation}
\delta \mathbf{x} = \frac{ \sum_i \gamma(\mathbf{x}_0, \mathbf{x}_c^i) \cdot \delta \mathbf{x}_c^i }{ \sum_i \gamma(\mathbf{x}_0, \mathbf{x}_c^i)  },
\end{equation}
where $i$ is the index of the three nearest control points. 

Since the position offset of a Gaussian is interpolated from nearby control points, by constraining the neighboring control points to have similar position offsets, the Gaussians among these control points are also constrained to move in the same direction. This ensures that the Gaussians can be optimized being distributed on a surface layer. \cref{fig:anchorpoint} provides an illustration of the control point design.

\emph{Discussion.} Note an alternative approach is to utilize another MLP to predict the position offsets for each control point. However, this approach produces blurry results (see 
\cref{fig:ablationanchorpoint} ``w/ MLP position offset"). This is because using an MLP to learn position offsets for all control points in a sequence can exceed the network's learning capacity, making it struggle to learn position offsets for each control point.

\subsection{Training and Testing}
\label{sec:trainandtest}

\noindent \textbf{Implementation Details.} We initialize $N=200K$ Gaussians on the template mesh, with each Gaussian assigned the skinning weights according to AnimatableGaussians~\cite{li2024animatable}. We also uniformly sample $F=300$ anchor points and $C=10K$ control points on the template mesh. The Gaussian neutral positions $\mathbf{x}_0$, anchor points $\mathbf{x}_a$, and control points $\mathbf{x}_c$ are fixed and not optimized once sampled on the mesh. The spatially distributed MLP has four layers. The pose vector used as MLP input does not include finger joints, as we believe that changes in fingers do not affect the overall appearance. The basis number $B$ is set to $15$. Each sequence is trained for 800K iterations. Our method is implemented using PyTorch, and we are able to achieve fast rendering speed without using acceleration techniques like CUDA or TensorRT, unlike previous method~\cite{zheng2023avatarrex}.

\noindent \textbf{Training.} During training, we simultaneously learn the spatially distributed MLPs, Gaussian neutral properties and property offset basis, as well as the neural position offsets and position offset basis for control points. We set the learning rate as $5\times10^{-4}$ for $\{ \mathbf{r}_0, \mathbf{s}_0, o_0, \mathbf{c}_0 \}$ , $1.6\times 10^{-4}$ for $\delta \mathbf{x}_{c0}$, and $5\times10^{-4}$ for spatially distributed MLPs. The learning rates for the basis are five times smaller. We also use learning rate decay in our implementation. 

For the loss functions, we use the L1 loss as in 3DGS~\cite{kerbl20233d} and LPIPS~\cite{zhang2018unreasonable} loss. The nearby control points are constrained to have similar position offsets $\delta \mathbf{x}_c$:
\begin{equation}
\mathcal{L}_{ctrl} = \sum_{i,j} \| \delta \mathbf{x}_c^i - \delta \mathbf{x}_c^j \|_2,
\end{equation}
where $i,j$ are the indices of nearby control points. We also limit the Gaussian scale to prevent the Gaussians being too large:
\begin{equation}
\mathcal{L}_{scale} = \sum_{i=1}^{N} \max(0.01, \mathbf{s}^i).
\end{equation}
The final loss is 
\begin{equation}
\mathcal{L} = \mathcal{L}_1 + 0.1 \mathcal{L}_{lpips} + 0.1 \mathcal{L}_{ctrl} + \mathcal{L}_{scale}.
\end{equation}

\noindent \textbf{Testing.} 
Following AnimatableGaussians~\cite{li2024animatable}, we use PCA to project novel poses to the space of training poses. Specially, all pose vectors of training poses form a matrix $X$. we perform PCA on $X$ and then project the novel pose to the linear space before using it as input. This strategy can make the model generalize better on the novel pose by fitting it within the space of the training poses.

\begin{figure}[tbp]
  \begin{center}
    \includegraphics[width=0.95\columnwidth]{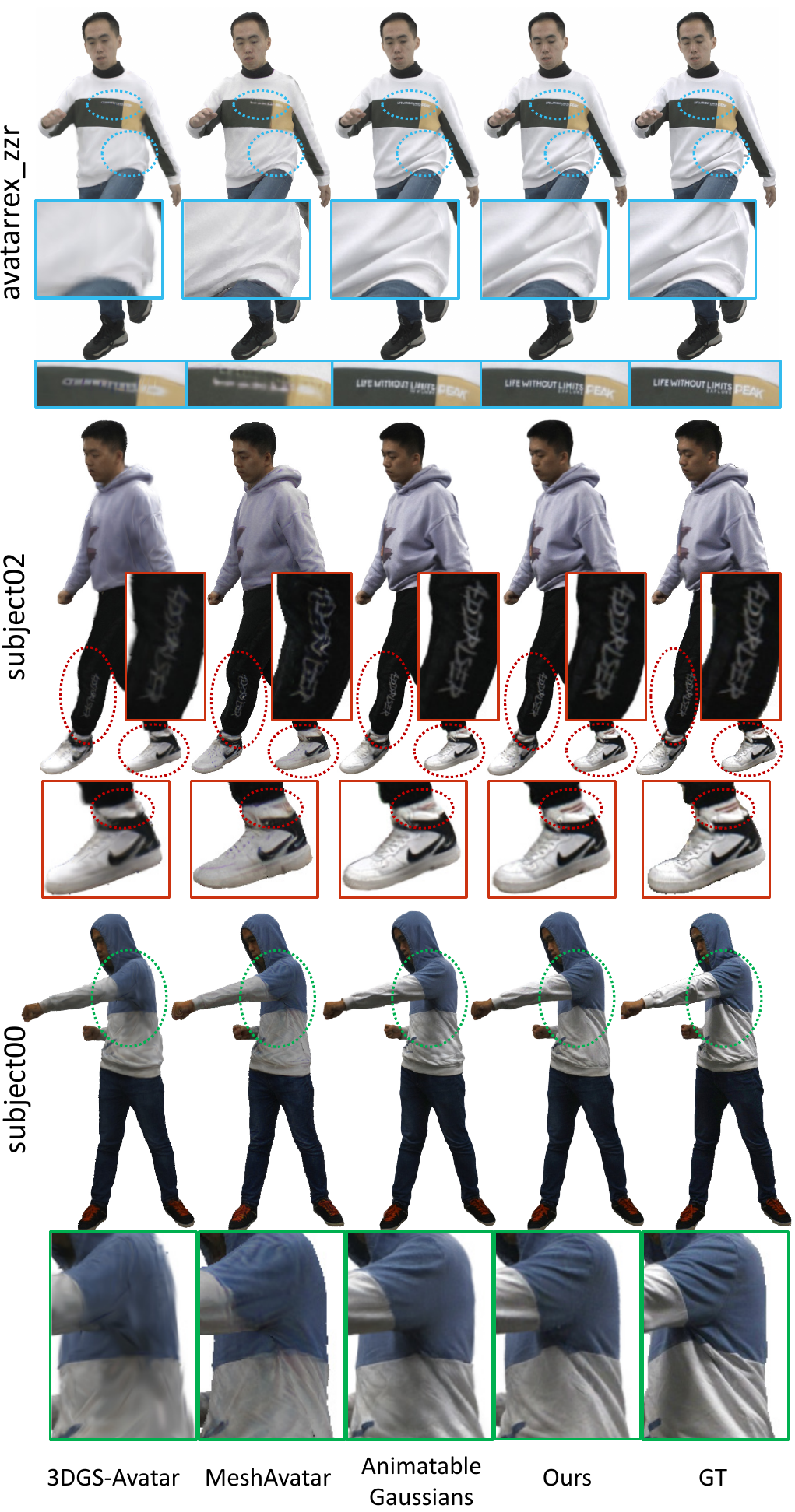}
  \end{center}
  \vspace{-5mm}
  \caption{Qualitative comparison with the state-of-the-art methods on training pose reconstruction (top two subjects) and novel pose synthesis (bottom subject). 
  }
  \label{fig:compare}
  \vspace{-5mm}
\end{figure}

\section{Experiments}

We conduct experiments on the following public datasets:

\noindent \textbf{AvatarRex~\cite{zheng2023avatarrex}.} This dataset contains full-body videos captured at 2K resolution from 16 views. We use 3 sequences and 14 views for our experiments. 

\noindent \textbf{THuman4.0~\cite{zheng2022structured}.} The dataset contains videos of people with rich wrinkles in their appearance. The videos are captured at 1K resolution. We use 3 sequences and 24 views for our experiments.

\noindent \textbf{ActorsHQ~\cite{isik2023humanrf}.} ActorsHQ is a high-quality dataset. We use 7 sequences and select 39 full-body views for training. We use 4x down-sampled images, with each image approximately 1K resolution.

Each sequence of the above datasets contains 1000 to 2000 frames. We also use the SMPL-X registrations provided by AnimatableGaussians~\cite{li2024animatable} for each sequence.

For quantitative experiments, we use Peak Signal-to-Noise Ratio (PSNR), Structural Similarity Index Measure (SSIM), Learned Perceptual Image Patch Similarity (LPIPS~\cite{zhang2018unreasonable}), and Frechet Inception Distance (FID~\cite{heusel2017gans}) as metrics. We calculate PSNR, SSIM, and FID on the whole image, and LPIPS  within the body bounding box. Training time and rendering speed are all evaluated on a single NVIDIA 3090 card. 

\begin{table}[tbp]
\centering
\resizebox{\columnwidth}{!}{
\begin{tabular}{l|cccc}
\toprule
Method              & PSNR $\uparrow$   & SSIM $\uparrow$  & LPIPS $\downarrow$  & FID $\downarrow$    \\ \hline
3DGS-Avatar~\cite{qian20243dgs}         & 28.9530 & 0.9741 & 0.0464 & 24.9938 \\
MeshAvatar~\cite{chen2024meshavatar}          & 29.3154 & 0.9731 & 0.0397 & 19.7409 \\
AnimatableGaussians~\cite{li2024animatable} & 31.2992 & 0.9831 & 0.0251 & 11.3421 \\
Ours                & \textbf{32.7456} & \textbf{0.9868} & \textbf{0.0226} & \textbf{10.1169} \\ \bottomrule
\end{tabular}
}
\vspace{-3mm}
\caption{Quantitative comparison with the state-of-the-art methods under training poses.}
\label{table:comparetrain}
\end{table}

\begin{table}[tbp]
\centering
\resizebox{\columnwidth}{!}{
\begin{tabular}{l|cccc}
\toprule
Method              & PSNR $\uparrow$   & SSIM $\uparrow$  & LPIPS $\downarrow$  & FID $\downarrow$    \\ \hline
PoseVocab~\cite{li2023posevocab}          & 26.3784 & 0.9707 & 0.0592 & 49.4541 \\
3DGS-Avatar~\cite{qian20243dgs}         & 27.5524 & 0.9737 & 0.0597 & 31.0979 \\
MeshAvatar~\cite{chen2024meshavatar}          & 27.4025 & 0.9717 & 0.0571 & 27.4278 \\
AnimatableGaussians~\cite{li2024animatable} & 28.1106 & 0.9741 & 0.0552 & 19.2324 \\
Ours                & \textbf{28.3263} & \textbf{0.9747} & \textbf{0.0537} & \textbf{19.0217} \\ \bottomrule
\end{tabular}
}
\vspace{-3mm}
\caption{Quantitative comparison with the state-of-the-art methods under novel poses.}
\label{table:comparenovel}
\end{table}

\begin{table}[tbp]
\centering
\resizebox{0.9\columnwidth}{!}{
\begin{tabular}{l|cc}
\toprule
Method              & Training Time (hours) $\downarrow$ & FPS $\uparrow$ \\ \hline
AnimatableGaussians~\cite{li2024animatable} & 100      & 10  \\
DEGAS~\cite{shao2024degas}               & 55       & 30  \\
Ours                & \textbf{17.5}       & \textbf{166} \\ \bottomrule
\end{tabular}
}
\vspace{-3mm}
\caption{Performance comparison. All methods are trained for 800K iterations. The training time and rendering fps under novel poses are recorded on a single NVIDIA 3090 card. For DEGAS~\cite{shao2024degas}, we directly adopt the data from the original paper.}
\label{table:speed}
\vspace{-3mm}
\end{table}

\subsection{Comparison}

\noindent \textbf{Quality.} We compare our method with 3DGS-Avatar~\cite{qian20243dgs}, MeshAvatar~\cite{chen2024meshavatar}, and AnimatableGaussians~\cite{li2024animatable}. We conduct experiments on sequence \textsf{avatarrex\_zzr} from AvatarReX dataset and sequence \textsf{subject00}, \textsf{subject02} from THuman4.0 dataset. \textsf{avatarrex\_zzr} and \textsf{subject02} are evaluated under training poses, while \textsf{subject00} is evaluated under novel poses. We present qualitative results in \cref{fig:compare}. In the cases of \textsf{avatarrex\_zzr} and \textsf{subject02}, 3DGS-Avatar and MeshAvatar struggle to learn fine wrinkles and textures. Although both our method and AnimatableGaussians can capture wrinkles similar to the ground truth, our method captures better details (e.g., the text ``LIFE WITHOUT LIMITS" on the chest of \textsf{avatarrex\_zzr} and the socks of \textsf{subject02}). In \textsf{subject00}, 3DGS-Avatar and MeshAvatar are not able to render high-quality appearances under novel poses, while our method and AnimatableGaussians can produce details like wrinkles. Therefore, our method surpasses other state-of-the-art methods like 3DGS-Avatar and MeshAvatar, and achieves comparable results to AnimatableGaussians, but with higher-fidelity details. For quantitative results, \cref{table:comparetrain} presents the metric comparison under training poses and \cref{table:comparenovel} under novel poses. Our method achieves the best results, demonstrating our method has strong learning ability in modeling human avatars and generalizes well to novel poses.

\noindent \textbf{Speed.} We also compare the training and rendering speed with AnimatableGaussians~\cite{li2024animatable} and DEGAS~\cite{shao2024degas}, which could render high-quality human avatars as well. Since DEGAS hasn't released its code, we directly use the performance data from its paper. \cref{table:speed} shows the training time and rendering speed. Our method uses less time to train and significantly outperforms other methods in rendering speed. This is because these methods use a large CNN to predict Gaussian properties, requiring a lot of time to go through the network (i.e., AnimatableGaussians takes 107ms to infer their network once). In contrast, our method takes only about 6ms to render a frame for 200K Gaussians, with 1.5ms for obtaining Gaussian coefficients, 3.3ms for combining the basis and LBS, and 1.1ms for rasterization, making the rendering process highly efficient.

\begin{figure}[t]
  \begin{center}
    \includegraphics[width=0.95\columnwidth]{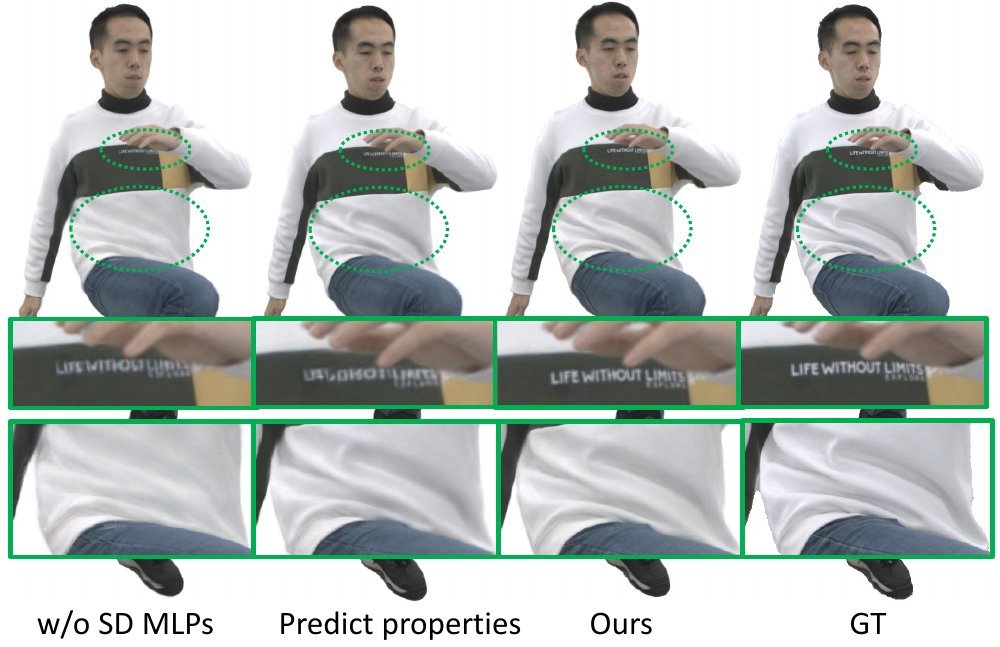}
  \end{center}
  \vspace{-5mm}
  \caption{Qualitative comparison of several design choices.
  }
  \label{fig:ablationnobs}
  \vspace{-2mm} 
\end{figure}

\begin{figure}[t]
  \begin{center}
    \includegraphics[width=0.95\columnwidth]{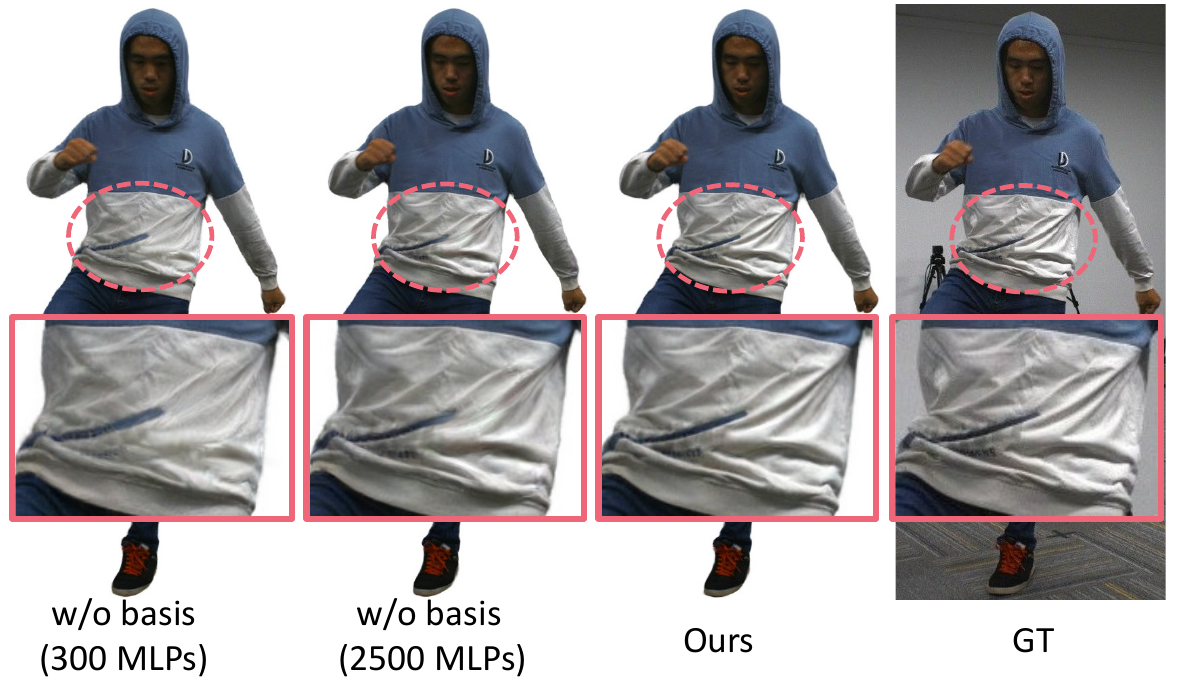}
  \end{center}
  \vspace{-5mm}
  \caption{Ablation study on Gaussian offset basis.
  }
  \label{fig:ablationinterprop}
  \vspace{-5mm} 
\end{figure}

\subsection{Ablation Study}

In this section, we evaluate several of our key designs. These designs impact the results in terms of quality or efficiency.

\noindent \textbf{Spatially Distributed MLPs.} Using multiple spatially distributed MLPs enhances the model's learning ability. For comparison, we use a single MLP to output the coefficient to combine the basis. \cref{table:ablation} ``w/o SD MLPs" shows that this approach greatly decreases the metrics, proving that a single MLP lacks sufficient learning capacity. \cref{fig:ablationnobs} ``w/o SD MLPs" also demonstrates that a single MLP is inadequate for capturing appearance details.

We also evaluate how the number of spatially distributed MLPs will influence the results. \cref{table:ablationmlpnum} shows the quality metrics and speed. To balance quality and efficiency, we choose to use $F=300$ MLPs.

\begin{figure}[tbp]
  \begin{center}
    \includegraphics[width=0.95\columnwidth]{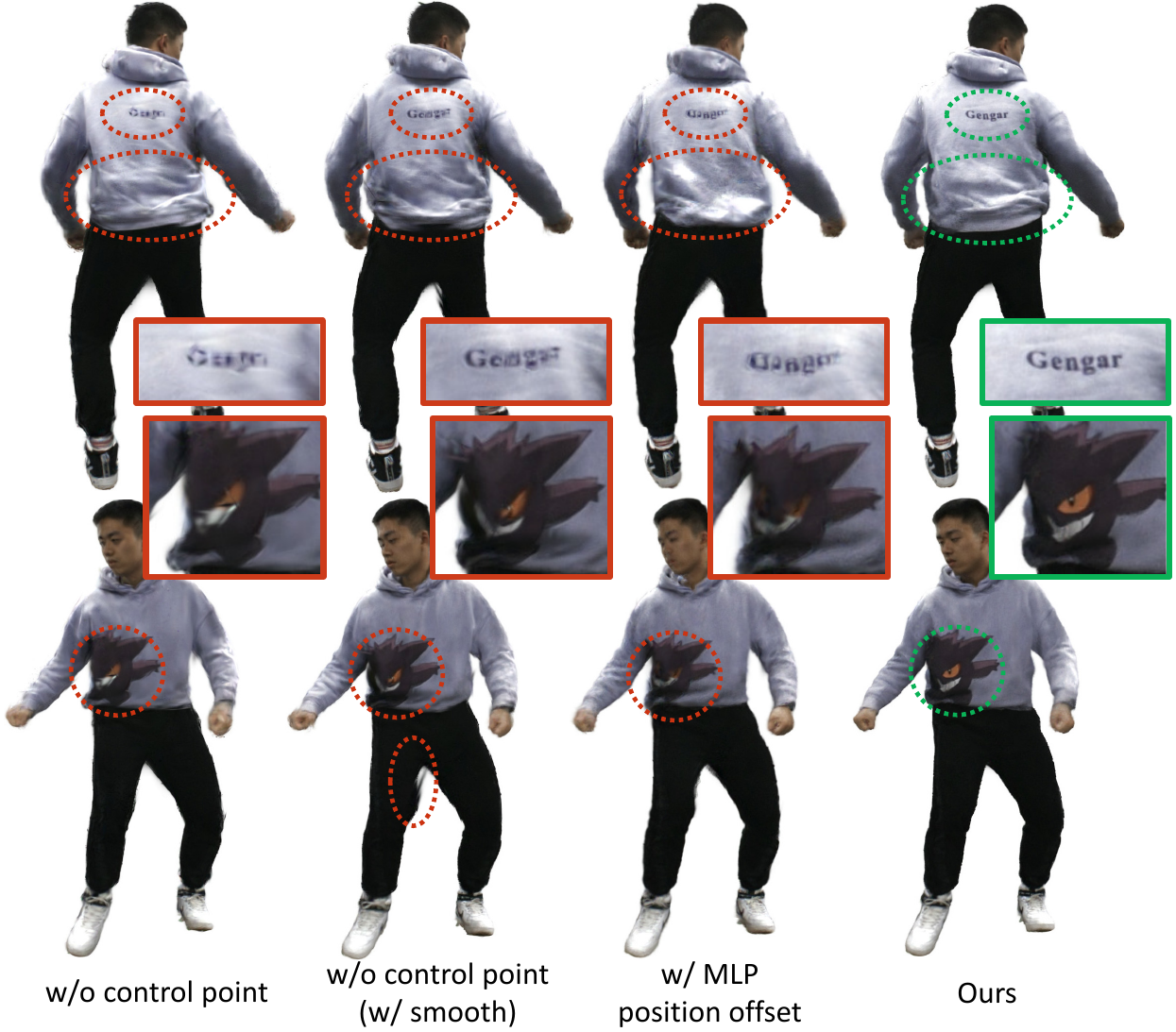}
  \end{center}
  \vspace{-5mm}
  \caption{Ablation study on control point.
  }
  \label{fig:ablationanchorpoint}
  \vspace{-2mm}
  
\end{figure}

\begin{table}[]
\centering
\resizebox{\columnwidth}{!}{
\begin{tabular}{l|cccccc}
\toprule
                  & PSNR $\uparrow$   & SSIM $\uparrow$  & LPIPS $\downarrow$  & FID $\downarrow$ & Training $\downarrow$ & FPS $\uparrow$  \\ \hline
w/o SD MLPs & 30.1941 & 0.9791 & 0.0314 & 15.4545 & 15.8 h & 182 \\
w/o basis (300 MLPs) & 31.8444 & 0.9832 & 0.0294 & 14.4292 & \textbf{15.7 h} & \textbf{205} \\
w/o basis (2500 MLPs) & 32.3912 & 0.9854 & 0.0244 & 11.0721 & 27.0 h & 115 \\ 
Predict properties & 31.9340 & 0.9834 & 0.0260 & 10.9757 & 24.2 h & 51 \\
w/o control point & 32.6780 & 0.9859 & 0.0241 & 10.6394 & 17.8 h & 161 \\
w/ MLP position offset & 31.8514 & 0.9816 & 0.0338 & 17.8664 & 18.4 h & 148 \\ \hline
Ours              & \textbf{32.7456} & \textbf{0.9868} & \textbf{0.0226} & \textbf{10.1169}  &{17.5 h} & {166}\\ \bottomrule
\end{tabular}
}
\vspace{-3mm}
\caption{Ablation study on design choices.}
\label{table:ablation}
\end{table}

\begin{table}[]
\centering
\resizebox{\columnwidth}{!}{
\begin{tabular}{l|cccccc}
\toprule
MLP number            & PSNR $\uparrow$   & SSIM $\uparrow$  & LPIPS $\downarrow$  & FID $\downarrow$ & Training $\downarrow$ & FPS $\uparrow$  \\ \hline
50 & 32.6625 & 0.9863 & 0.0237 & 11.1197 & 16.6 h & 173 \\
300 (Ours) & 32.7456 & 0.9868 & 0.0226 & 10.1169  & 17.5 h & 166\\ 
800 & 32.7313 & 0.9866 & 0.0224 & 9.8885 & 20.8 h & 149 \\\bottomrule
\end{tabular}
}
\vspace{-3mm}
\caption{Quantitative comparison of different number of spatially distributed MLPs.}
\label{table:ablationmlpnum}
\end{table}

\begin{table}[]
\centering
\resizebox{\columnwidth}{!}{
\begin{tabular}{l|cccccc}
\toprule
Basis number            & PSNR $\uparrow$   & SSIM $\uparrow$  & LPIPS $\downarrow$  & FID $\downarrow$ & Training $\downarrow$ & FPS $\uparrow$  \\ \hline
5 & 32.3523 & 0.9854 & 0.0248 & 11.5850 & 16.5 h & 175 \\
15 (Ours) & 32.7456 & 0.9868 & 0.0226 & 10.1169  & 17.5 h & 166\\
40 & 32.7154 & 0.9866 & 0.0221 & 10.0522 & 20.4 h & 156 \\ 
40 (large MLP) & 32.7010 & 0.9869 & 0.0220 & 9.9707 & 24.6 h & 140 \\ \bottomrule
\end{tabular}
}
\vspace{-3mm}
\caption{Quantitative comparison of different number of the basis.}
\label{table:ablationbasisnum}
\vspace{-3mm}
\end{table}

\noindent \textbf{Gaussian Offset Basis.} We demonstrate that high-frequency details are difficult to recover without Gaussian offset basis. To prove this, we let the spatially distributed MLPs output Gaussian property offsets, and perform interpolation on the Gaussian property offsets. We experiment this idea using 300 and 2500 MLPs. \cref{fig:ablationinterprop} ``w/o basis (300 MLPs)" and ``w/o basis (2500 MLPs)" cannot recover all the fine details. In comparison, our results are almost identical to the ground truth. For quantitative comparison, \cref{table:ablation} ``w/o basis (300 MLPs)" and ``w/o basis (2500 MLPs)" show a decline in metrics compared with ours. Also, increasing the number of MLPs to 2500 can also reduce rendering speed, as shown in \cref{table:ablation} ``w/o basis (2500 MLPs)".  

We also shows that other possible designs are not as good as our basis combination. For example, following \citet{ye2023animatable}, we assign a learnable vector to each Gaussian. Each Gaussian finds its nearest spatially distributed MLP and the MLP takes the learnable vector and pose vector as input and directly predicts Gaussian property offsets. \cref{fig:ablationnobs} ``Predict properties" shows the results, indicating that this design still fails to capture very high-frequency details (such as the text on the chest). Additionally, because the MLPs take the pose and learnable vector as input, they need to be inferred many times, thus the training time increases and rendering speed is significantly reduced, as shown in \cref{table:ablation} ``Predict properties".

\begin{figure}[tbp]
  \begin{center}
    \includegraphics[width=0.9\columnwidth]{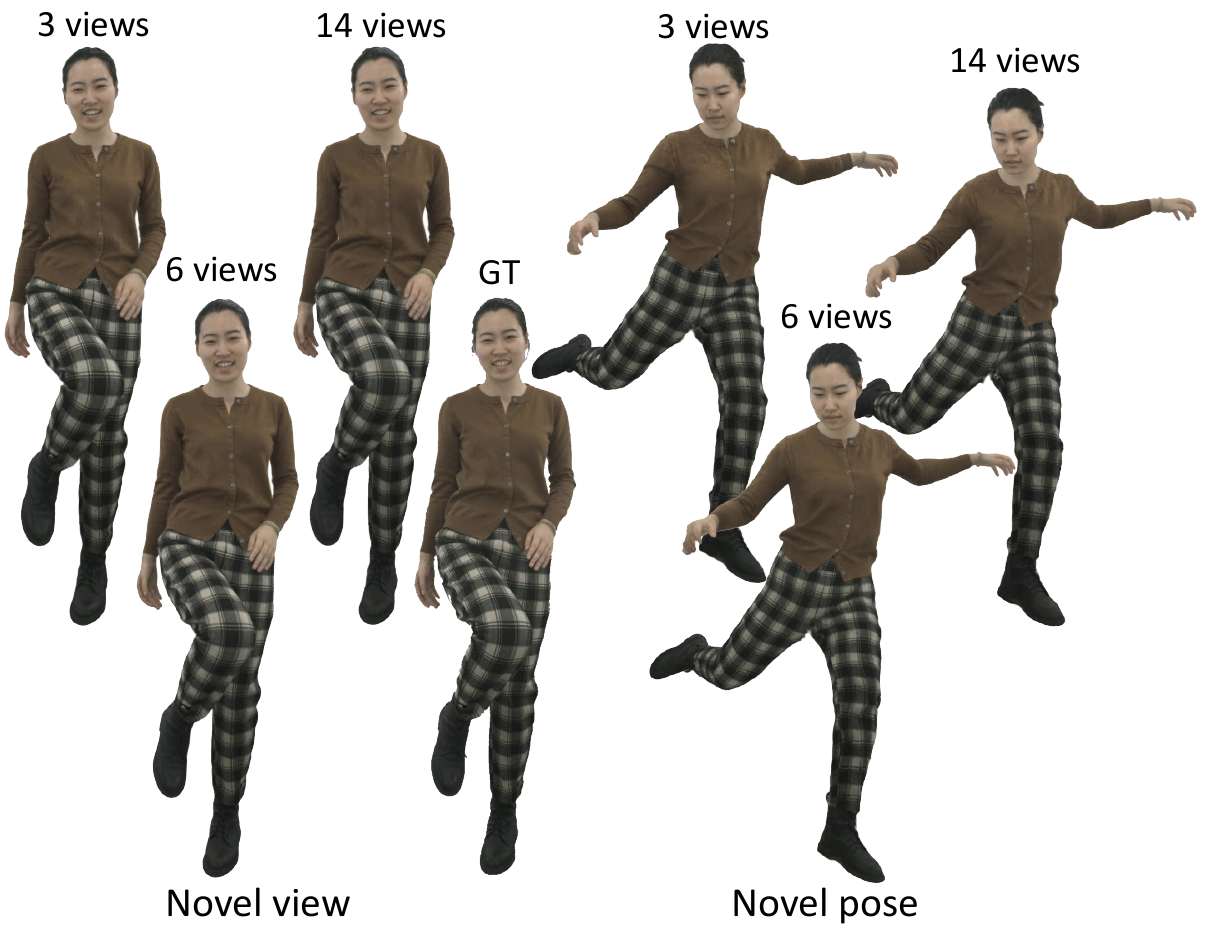}
  \end{center}
  \vspace{-5mm}
  \caption{Results of different view number.
  }
  \label{fig:sparseview}
  \vspace{-5mm}
\end{figure}

\begin{figure*}[t]
  \begin{center}
    \includegraphics[width=0.98\textwidth]{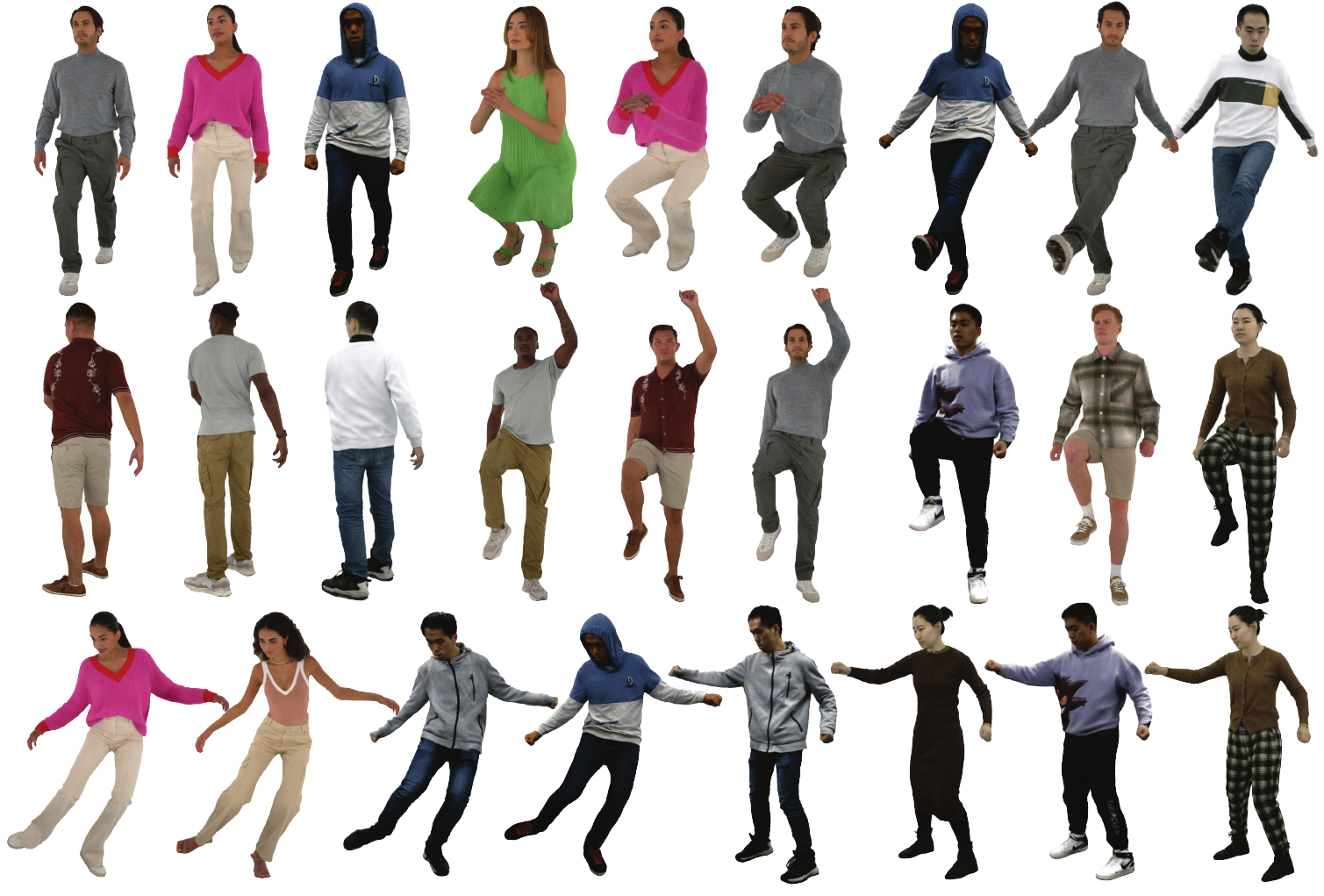}
  \end{center}
  \vspace{-5mm}
  \caption{Our method achieves high-quality human avatar reconstruction and animation under novel poses.
  }
  \label{fig:gallery}
  \vspace{-3mm}
\end{figure*}

We also verify the suitable number of basis. \cref{table:ablationbasisnum} shows quantitative results for 5, 15, and 40 bases. The results of 15 and 40 bases are comparable with tiny gaps. A reasonable guess is that the linear sub-space of the Gaussian property offsets space can already be well supported with 15 bases. Increasing to 40 bases does not necessarily improve the representation capability. To further validate the guess, we use larger MLPs to learn 40 coefficients for 40 bases, in case the performance is limited by the MLP capacity, but the results are still comparable with those of 15 bases with slight differences (see \cref{table:ablationbasisnum}). Empirically, we choose $B=15$ as our selection.

\noindent \textbf{Control Point.} Without control points, Gaussians can move freely, resulting in suboptimal results when animated under novel pose. \cref{fig:ablationanchorpoint} ``w/o control point" shows qualitative result. Even with additional local smoothness constraints (\cref{fig:ablationanchorpoint} ``w/o control point (w/ smooth)" ), Gaussians still cannot be well-constrained to the surface, causing details such as text and textures to become corrupted under novel pose. We also demonstrate that utilizing another MLP to predict the position offsets for each control point is still insufficient to render good results (\cref{fig:ablationanchorpoint} ``w/ MLP position offset" ). This is because using an MLP to learn position offsets for all control points in a sequence may still exceed one MLP's learning capacity. \cref{table:ablation} ``w/ MLP position offset" also shows decline in metrics.

\noindent \textbf{Sparse View.} Our method can be trained under sparse viewpoints. We evaluate it using sequence \textsf{avatarrex\_lbn2} from AvatarRex dataset, and present results in \cref{fig:sparseview} using 3, 6 and 14 viewpoints. Our method achieves high-fidelity results even with only three viewpoints for training and can still be animated under novel poses.

\subsection{More Results}

We show more results in \cref{fig:gallery}. The presented results are animated under novel poses. Our method achieves high-quality human avatar reconstruction and animation. We also provide a viewer that allows users to interactively animate the reconstructed human avatar. Please refer to the supplementary video for more visual results. We note that fps in the viewer is slightly lower due to additional data transmission overhead.

\section{Conclusion and Limitation}

In this paper, we propose a method capable of modeling high-fidelity human avatars with high-frequency details while the rendering speed is very fast. We use the spatially distributed MLPs to infer the coefficients for the Gaussian offset basis. The smoothly interpolated coefficients combined with freely learned basis can produce distinctly different Gaussian property offsets, allowing the ability to learn high-frequency details. We also use control points to constrain the Gaussians to be distributed on a surface layer without moving inside the body. Experiments demonstrate that our method surpasses previous state-of-the-art methods both in reconstruction fidelity and rendering performance. Currently our avatar appearance is conditioned on pose and cannot model other complex cloth dynamics such as long skirt swaying in the wind.  Modeling clothes as a separate layer and incorporating simulation could potentially improve the applicability of our model. Reconstructing human avatars with high-fidelity pose-dependent appearances from monocular videos is another direction worth exploring. Our method still relies on multi-view capture, pose estimation, and template mesh extraction, which makes the pipeline quite heavy. In future we plan to use fewer RGBD cameras to reduce the complexity of pipeline, as the difficulties of skeleton estimation and mesh reconstruction can be largely reduced with the depth information.

\clearpage

\section*{Acknowledgment}

The authors would like to thank the reviewers for their insightful comments. This work is supported by the National Key Research and Development Program of China (No.2022YFF0902302), NSF China (No. 62322209 and No. 62421003), the gift from Adobe Research, the XPLORER PRIZE, and the 100 Talents Program of Zhejiang University. The source code is available at \textcolor{magenta}{ \href{https://gapszju.github.io/mmlphuman}{https://gapszju.github.io/mmlphuman}}.
{
    \small
    \bibliographystyle{ieeenat_fullname}
    \bibliography{main}
}
\clearpage
\setcounter{page}{1}
\setcounter{figure}{0}  
\setcounter{section}{0}
\maketitlesupplementary

\section{More Implementation Details}

\noindent \textbf{Template Mesh.} For the canonical template mesh used in our method, we use SMPL-X~\cite{pavlakos2019expressive} mesh as the template mesh for avatars wearing tight clothing, and follow  AnimatableGaussians~\cite{li2024animatable} to obtain template mesh for avatars wearing loose clothing.

\noindent \textbf{Spatially Distributed MLPs.} The spatially distributed MLP contains four hidden layers with 512, 256, 256, 256 neurons respectively. Each spatially distributed MLP takes only pose vector as input and outputs coefficients of length $2B=30$. Half of the coefficients are interpolated by Gaussians to obtain Gaussian coefficients, while the other half are interpolated by control points to obtain control point coefficients. 

For the implementation details, we use the new \textsf{vmap} function in Pytorch 2.0 for the parallel processing of multiple spatially distributed MLPs. The \textsf{vmap} function allows running models (e.g., MLPs and CNNs) with the same architecture in parallel. Specifically, we convert the MLP to a function (line 6 in Code~\ref{samplecode}), and then \textsf{vmap} runs the function with the batched weights and tensors as input (line 8 in Code~\ref{samplecode}). Please refer to the official document \textsf{pytorch.org/tutorials/intermediate/ensembling.html} for more details.

We also compare the forward time of \textsf{vmap} with group 1D convolution, which was used in SLRF~\cite{zheng2022structured} to run multiple MLPs in parallel. We run 1000 iterations and report the average time. The \textsf{vmap} implementation takes 0.52ms, and is much faster than the group convolution (3.70ms).

\noindent \textbf{Training Details.} At the beginning of training, we only optimize Gaussian neutral properties and neural position offsets. After 2K iterations, we optimize property offset basis and position offset basis, as well as spatially distributed MLPs. We also optimize only the zero-order component of the SH coefficients at the beginning. The first-order SH coefficients are optimized after 250K iterations.

\setcounter{table}{0}  

\newfloat{code}{bp}{loc}[section]
\floatname{code}{Code}

\begin{table}[bp]
\vspace{-7mm}
\captionsetup{labelformat=default,labelsep=period,name=Code}
\begin{center}
    
\begin{lstlisting}[language=Python]
from torch.func import vmap, functional_call, stack_module_state
models = [MLP(in=63,out=30) for _ in range(300)]
weights, _ = stack_module_state(models)
base_model = MLP(in=63,out=30).to("meta")
def fmodel(param, data):
  return functional_call(base_model, param, data)
input_tensor = torch.tile(pose_vector, [300, 1])
output_tensor = vmap(fmodel)(weights, input_tensor)
# output tensor shape: (300, 30)
\end{lstlisting}
\end{center}
\vspace{-8mm}
\caption{Sample code for processing multiple MLPs in parallel.}
\label{samplecode}
\vspace{-9mm}
\end{table}

\setcounter{table}{0}  

\section{Experiment Details}

In the main paper, the quantitative results in Tab. 1, Tab. 4, Tab. 5, and Tab. 6 are evaluated on \textsf{avatarrex\_zzr} sequence from AvatarRex~\cite{zheng2023avatarrex} dataset. We use the first 2000 frames for training and calculate metrics on the first 500 frames from ``22010710" camera view. The quantitative results in Tab. 2 of the main paper are evaluated on \textsf{subject00} sequence from THuman4.0~\cite{zheng2022structured} dataset. We use the first 2000 frames for training and calculate metrics on the remaining 500 frames from ``cam18" camera view.

\section{Ablation Study on Gaussian Number}

\begin{figure}[t]
  \begin{center}
    \includegraphics[width=0.95\columnwidth]{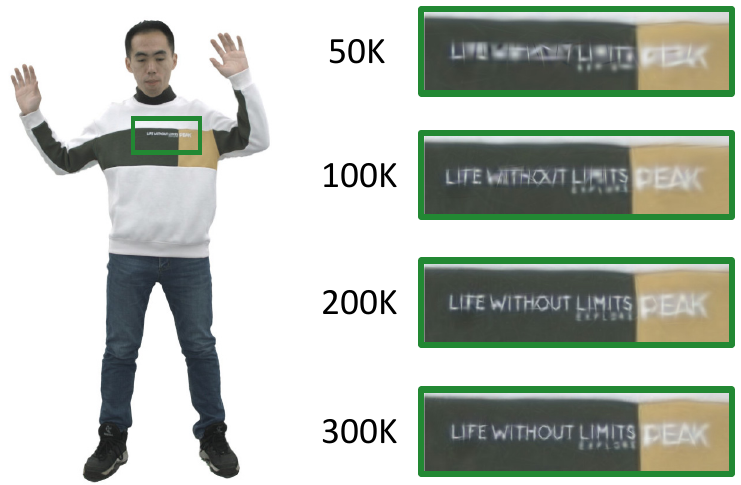}
  \end{center}
  \vspace{-5mm}
  \caption{Qualitative comparison of different number of Guassians.
  }
  \label{fig:ablationgsnum}
\end{figure}

We conduct experiments on the number of Gaussians used in our method. Quantitative experiments are shown in \cref{table:ablationgsnum}. Although reducing the number of Gaussians can greatly improve rendering speed, we find that fewer Gaussians make it more difficult to accurately capture details, as shown in \cref{fig:ablationgsnum}. Therefore, we empirically choose 200K Gaussians in our method.

\begin{table}[]
\centering
\resizebox{\columnwidth}{!}{
\begin{tabular}{l|cccccc}
\toprule
GS number          & PSNR $\uparrow$   & SSIM $\uparrow$  & LPIPS $\downarrow$  & FID $\downarrow$ & Training $\downarrow$ & FPS $\uparrow$  \\ \hline
50K & 32.6020 & 0.9858 & 0.0244 & 10.7696 & 15.0 h & 332 \\
100K & 32.6833 & 0.9863 & 0.0232 & 10.3303  & 15.8 h & 246\\ 
200K & 32.7456 & 0.9868 & 0.0226 & 10.1169 & 17.5 h & 166 \\
300K & 32.7626 & 0.9869 & 0.0224 & 10.0621 & 19.0 h & 129 \\\bottomrule
\end{tabular}
}
\caption{Quantitative comparison of different number of Gaussians.}
\label{table:ablationgsnum}
\end{table}

\section{Ablation Study on PCA Components}

\begin{figure}[htbp]

  \begin{center}
    \includegraphics[width=0.99\columnwidth]{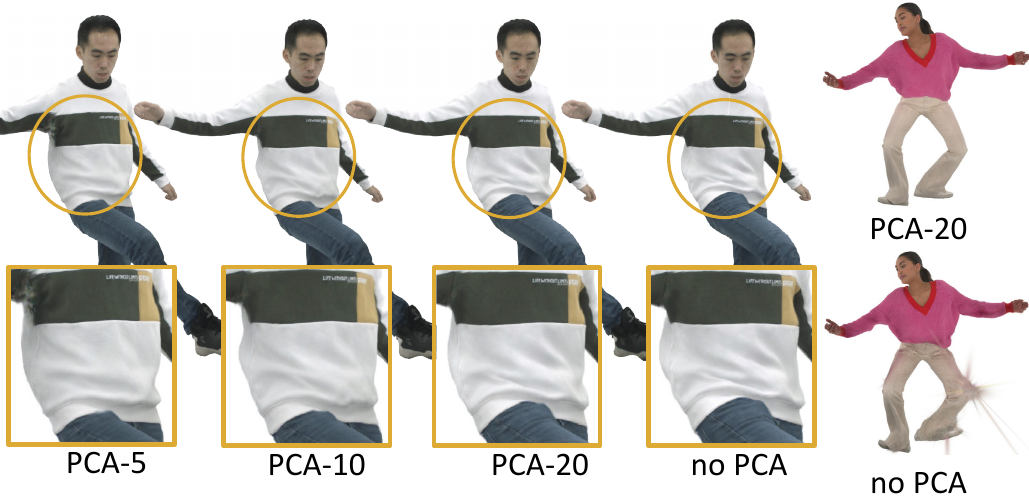}
  \end{center}
  \vspace{-5mm}
  \caption{Ablation study on PCA components.
  }
  \label{fig:ablationpca}
  
\end{figure}

We use 20 PCA components during testing. \cref{fig:ablationpca} shows the results of using different numbers of components and no PCA. The left results show that using fewer PCA components yields fewer details. The right results show that artifacts can appear when no PCA is used.

\end{document}